\pdfoutput=1
\documentclass[3p,times,twocolumn]{elsarticle}
\usepackage{ecrc}

\volume{00}
\firstpage{1}

\journalname{Nuclear Physics B Proceedings Supplement}

\jid{nppp}

\jnltitlelogo{Nuclear Physics B Proceedings Supplement}

\usepackage{lineno}

\usepackage{amssymb}
\usepackage[figuresright]{rotating}
\usepackage{graphicx}
\usepackage{xspace}

\newcommand{\snn}{\ensuremath{\sqrt{s_{NN}}}\xspace}
\newcommand{\Ups}{\ensuremath{\Upsilon}\xspace}
\newcommand{\Jpsi}{\ensuremath{\mathrm{J}/\psi}\xspace}
\newcommand{\pT}{\ensuremath{p_\mathrm{T}}\xspace}

\newcommand{\Raa}{\ensuremath{R_\mathrm{AA}}\xspace}
\newcommand{\Rda}{\ensuremath{R_\mathrm{dAu}}\xspace}
\newcommand{\Npart}{\ensuremath{N_\mathrm{part}}\xspace}
\newcommand{\Ncoll}{\ensuremath{N_\mathrm{coll}}\xspace}

\newcommand{\Bee}{B_\mathrm{ee}\xspace}
\newcommand{\ee}{\mathrm{e}^{-}\mathrm{e}^{+}\xspace}

\begin{document}

\begin{frontmatter}

\dochead{}

\title{Bottomonium production in heavy-ion collisions at STAR}

\author{R\'obert V\'ertesi\fnref{emil} (for the STAR Collaboration)}
\runauth{R. V\'ertesi for STAR}
\fntext[emil]{e-mail: \tt vertesi.robert@wigner.mta.hu}
\address{Nuclear Physics Institute of the Academy of Sciences of the Czech Republic, \\ \v{R}e\v{z} 130, 25068 Husinec-\v{R}e\v{z}, Czech Republic}
\address{Wigner Research Centre for Physics of the Hungarian Academy of Sciences RMI,\\ H-1525 Budapest 114, P.O.Box 49, Hungary}

\begin{abstract}
Bottomonium measurements provide unique insight into hot and cold nuclear matter effects present in the medium that is formed in high-energy heavy-ion collisions. 
Recent STAR results show that in $\snn = 200$ GeV central Au+Au collisions the \Ups{}(1S) state is suppressed more than the case that if only cold nuclear matter effects were present, and the excited state yields are consistent with a complete suppression. 
In 2012, STAR also collected 263.4 $\mu$b$^{-1}$ high-energy-electron triggered data in U+U collisions at $\snn = 193$ GeV. 
Central U+U collisions, with an estimated 20\% higher energy density than that in central Au+Au data, extend the \Ups{}(1S+2S+3S) and \Ups{}(1S) nuclear modification trends observed in Au+Au towards higher number of participant nucleons, and confirm the suppression of the \Ups{}(1S) state. 
We see a hint with 1.8 $\sigma$ significance that the \Ups{}(2S+3S) excited states are not completely suppressed in U+U collisions. These data support the sequential in-medium quarkonium dissociation picture and favor models with a strong $q\bar{q}$ binding.
\end{abstract}

\begin{keyword}
Brookhaven RHIC Coll \sep quarkonium: heavy \sep quarkonium: production \sep quark gluon: plasma
\end{keyword}

\end{frontmatter}


\section{Introduction}
\label{sec:intro}

It has long been suggested that, due to the screening of the heavy quark potential, the yield of heavy quarkonia is suppressed in heavy-ion collisions compared to expectations from p+p collisions. Charmonium suppression was, in fact, anticipated as a key signature of quark-gluon plasma (QGP) formation~\cite{Matsui:1986dk}.
Moreover, it is expected that quarkonium states follow a sequential suppression pattern, where states with lower binding energies dissociate ("melt") at lower temperatures than states with higher binding energies, thus quarkonium measurements may serve as a thermometer of the medium~\cite{Mocsy:2007jz}. 
Despite early expectations, it has been found that the energy dependence of $J/\psi$ suppression is rather weak from SPS to top RHIC energies~\cite{Alessandro:2004ap,Adare:2006ns}. 
An explanation to this is that later recombination (coalescence) of $c\bar{c}$ pairs gives a sizeable contribution to the \Jpsi yield which compensates for melting. Moreover, d+Au measurements proved that the cold nuclear matter (CNM) effects also play an important role~\cite{Adare:2010fn}. 
Feed-down from heavier states such as $\chi_c$, $\psi^\prime$ and $B$ mesons to \Jpsi also contribute to the measured yields. 
Bottomonia are expected to be less affected by many of these effects than charmonia, as co-mover absorbtion and recombination are predicted to be negligible at RHIC energies~\cite{Lin:2000ke}. Therefore it provides a cleaner probe for QGP effects. Nevertheless, feed-down such as $\chi_b$, \Ups{}(2S) and \Ups{}(3S) decaying to \Ups{}(1S) is still present in the case of bottomonia.

\section{Upsilon measurements at STAR}

The STAR experiment at RHIC comprises several
subsystems that provide full azimuthal coverage at
mid-rapidity ($|\eta|${}$<$1). A detailed description of the STAR
detector can be found in Ref.~\cite{Ackermann:2002ad}.
The current measurements reconstruct \Ups states through the  dielectron decay channel. 

We quantify the modification of \Ups production in the nuclear matter produced in A+A collisions with the nuclear modification factor, 
\[
\Raa^\Upsilon = \frac{\sigma^{inel}_{pp}}{\sigma^{inel}_{AA}}
\frac{1}{\langle\Ncoll\rangle}
\frac{\Bee\times d\sigma^{AA}_\Ups / dy}{\Bee\times d\sigma^{pp}_\Ups / dy}\ ,
\]
where $\sigma^{inel}_{AA(pp)}$ is the total
inelastic cross section of the A+A (p+p) collisions,
$d\sigma^{AA(pp)}_\Ups /dy$ denotes the \Ups
production cross section in A+A (p+p) collisions, and $\Bee \approx 2.4\%$ is the branching ratio of the $\Upsilon\rightarrow\ee$ processes. A similarly defined \Rda is used for describing nuclear modification in d+Au collisions.
Measurements in p+p collisions are essential not only as a baseline for nuclear modification, but also as a benchmark for theoretical calculations. NLO pQCD calculations with the color evaporation model~\cite{Vogt:2012fba} describe the \Ups measurements at RHIC within their uncertainties~\cite{Adamczyk:2013poh}.

\begin{figure}[]
\centering
\includegraphics[height=60mm]{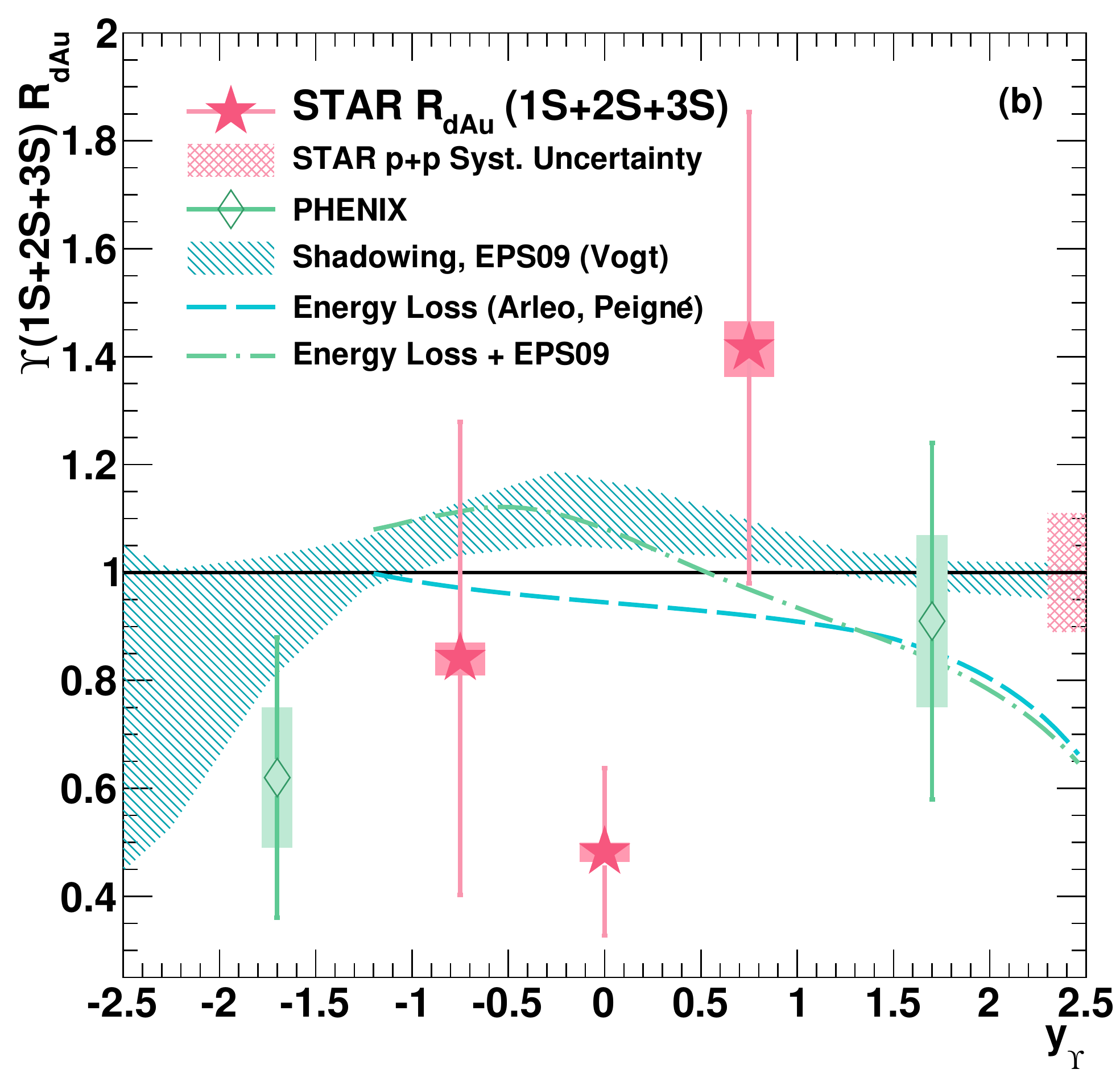}%
\vspace{-3mm}
\caption{\Rda versus $y$ of \Ups{} mesons~\cite{Adamczyk:2013poh} compared to theoretical calculations~\cite{Vogt:2012fba,Arleo:2012rs}.}
\label{fig:StarDauRaa}
\end{figure}

Fig.~\ref{fig:StarDauRaa} shows the \Rda versus rapidity measured by STAR and PHENIX~\cite{Adare:2012bv}. The models shown in the figure consider different CNM effects, and they predict a rather small nuclear modification in mid-rapidity d+Au collisions at $\snn{}=200$ GeV~\cite{Vogt:2012fba,Arleo:2012rs}. Interestingly, STAR measurements show a strongly suppressed \Ups production at mid-rapidity ($|y|<0.5$), although the uncertainties are relatively high~\cite{Adamczyk:2013poh}. This is a warning that CNM effects may also play an important role in nucleus-nucleus collisions, or that hot nuclear matter effects may also be present in d(p)+Au collisions. Analysis of the high-statistics p+Au data collected in 2015 is essential to clarify this question. 

Most of the heavy-ion collisions recorded at STAR are Au+Au data taken at $\snn{}=200$ GeV. Besides that, collisions of non-spherical Uranium nuclei are of particular interest; Central U+U data at $\sqrt{s_{NN}}$=193 GeV are estimated to have a 20\% higher average energy density than that of central Au+Au data at $\sqrt{s_{NN}}$=200 GeV~\cite{Kikola:2011zz}, thus allowing for further tests of the sequential suppression hypothesis. In the followings we summarize STAR bottomonium measurements in Au+Au collisions at $\snn{}=200$ GeV~\cite{Adamczyk:2013poh} as well as preliminary results in U+U collisions at $\snn=193$ GeV.

\section{Reconstruction and analysis of the A+A data}

Events of heavy-ion collisions were selected based on a ``high tower'' trigger that fired in each event where an energetic hit (approximately 4.2 GeV depending on the particular tower) was present in the Barrel Electromagnetic Calorimeter (BEMC). A total of 1.08 nb$^{-1}$ integrated luminosity was used in the case of Au+Au collisons, and 263.4 $\mu$b $^{-1}$ in the U+U measurement.
Momentum measurement and electron identification based on the specific energy loss $\mathrm{d}E/dx$ were done in the Time Projection Chamber (TPC).
Further electron identification was applied using the BEMC. 
The position of the reconstructed tracks were extrapolated towards the BEMC surface and matched to hits in BEMC towers.
Three adjacent towers around a BEMC hit that contained the highest energy deposits were combined into clusters in order to get a better estimate of the total energy deposit. 
Electron candidates were then required to have a cluster energy close to its momentum. Since electromagnetic showers are generally compact, most of the cluster energy was required to be in a single tower. The particular cut values differ in the two analyses; because of the scarcity of statistics in the U+U case, tighter cuts had to be applied in order to reduce the background.
\begin{figure}[]
\centering
\includegraphics[trim={0 0 0 30},width=0.5\linewidth,height=0.48\linewidth]{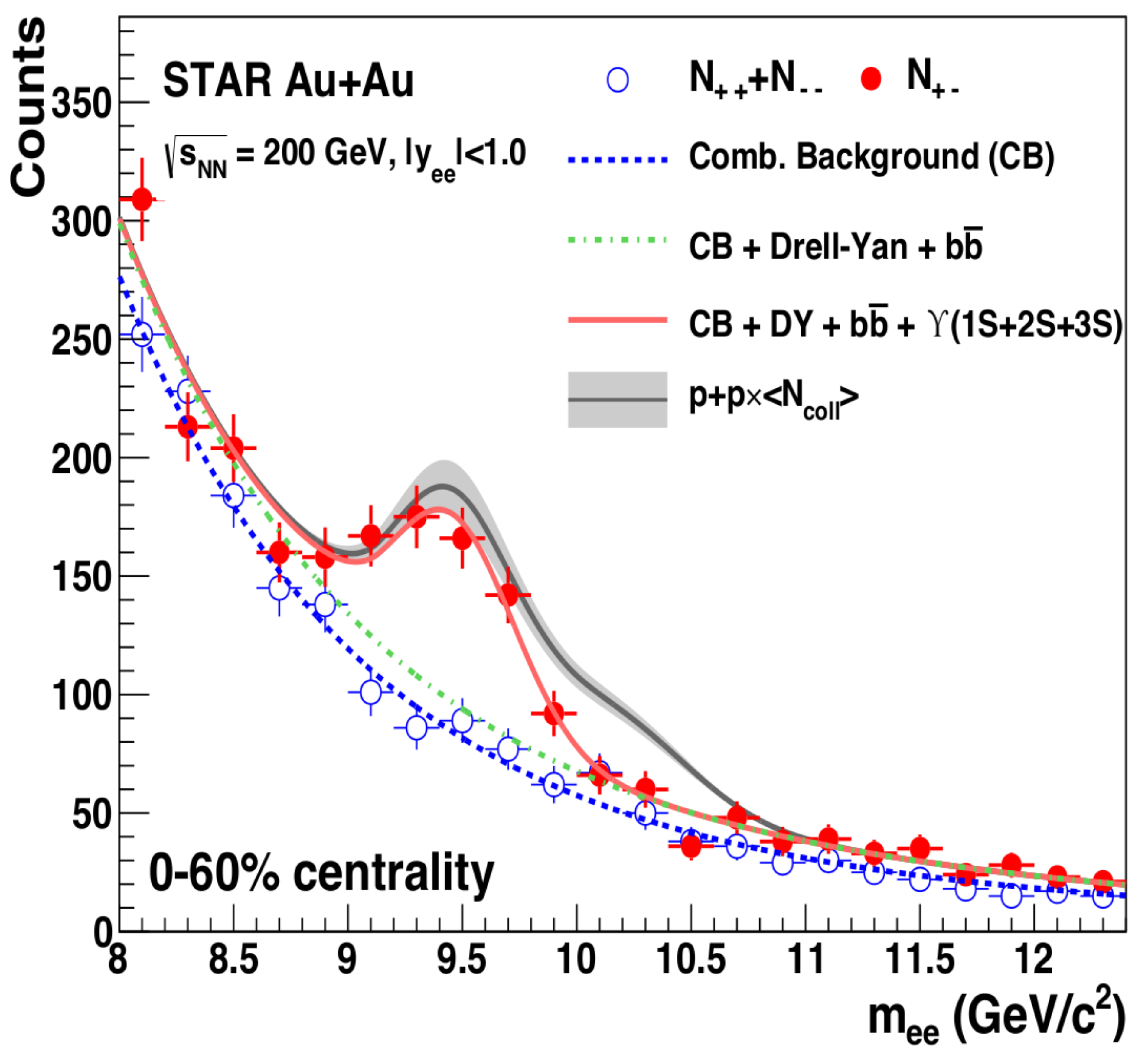}%
\includegraphics[width=0.5\linewidth,height=0.5\linewidth]{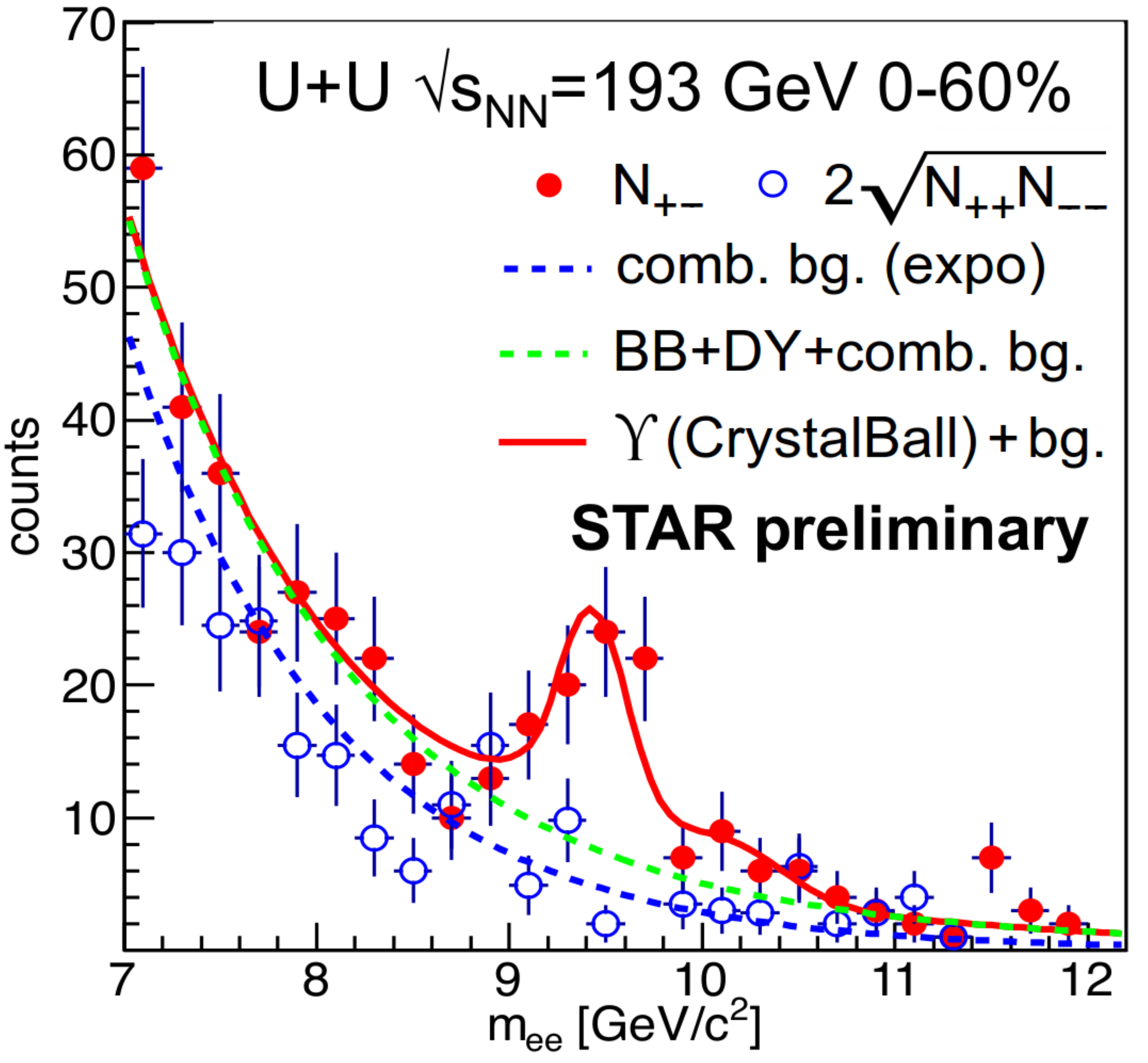}
\vspace{-5mm}
\caption{Invariant mass distributions of the \Ups{}(1S+2S+3S) candidates after selection cuts in 0-60\% centrality Au+Au data at $\sqrt{s_{NN}}$=200 GeV~\cite{Adamczyk:2013poh} (left), as well as U+U data at 193 GeV (right).}
\label{fig:UpsInvMass}
\end{figure}

The \Ups{}(1S+2S+3S) candidates were reconstructed as unlike-sign pairs of an electron that fired the trigger with another electron in the same event. Like-sign pairs were used to reconstruct the combinatorial background.
In the peak region there is also a significant
contribution from Drell-Yan and open $b\bar{b}$ processes. Templates of
the $\Ups(nS)$ peaks and the Drell-Yan contributions obtained from
simulations, and the $b\bar{b}$ contribution from pQCD model calculations were fitted simultaneously to determine their relative contributions. 

Fig.~\ref{fig:UpsInvMass} shows the reconstructed invariant mass distributions of unlike- and like-sign pairs in heavy-ion collisions, together with the peak and the combinatorial and correlated background fits.
The \Ups yield was determined using bin-counting after the subtraction of the fitted backgrounds. 

\section{Results}

\begin{figure}[h]
\centering
\vspace{-3mm}
\includegraphics[width=60mm,height=60mm]{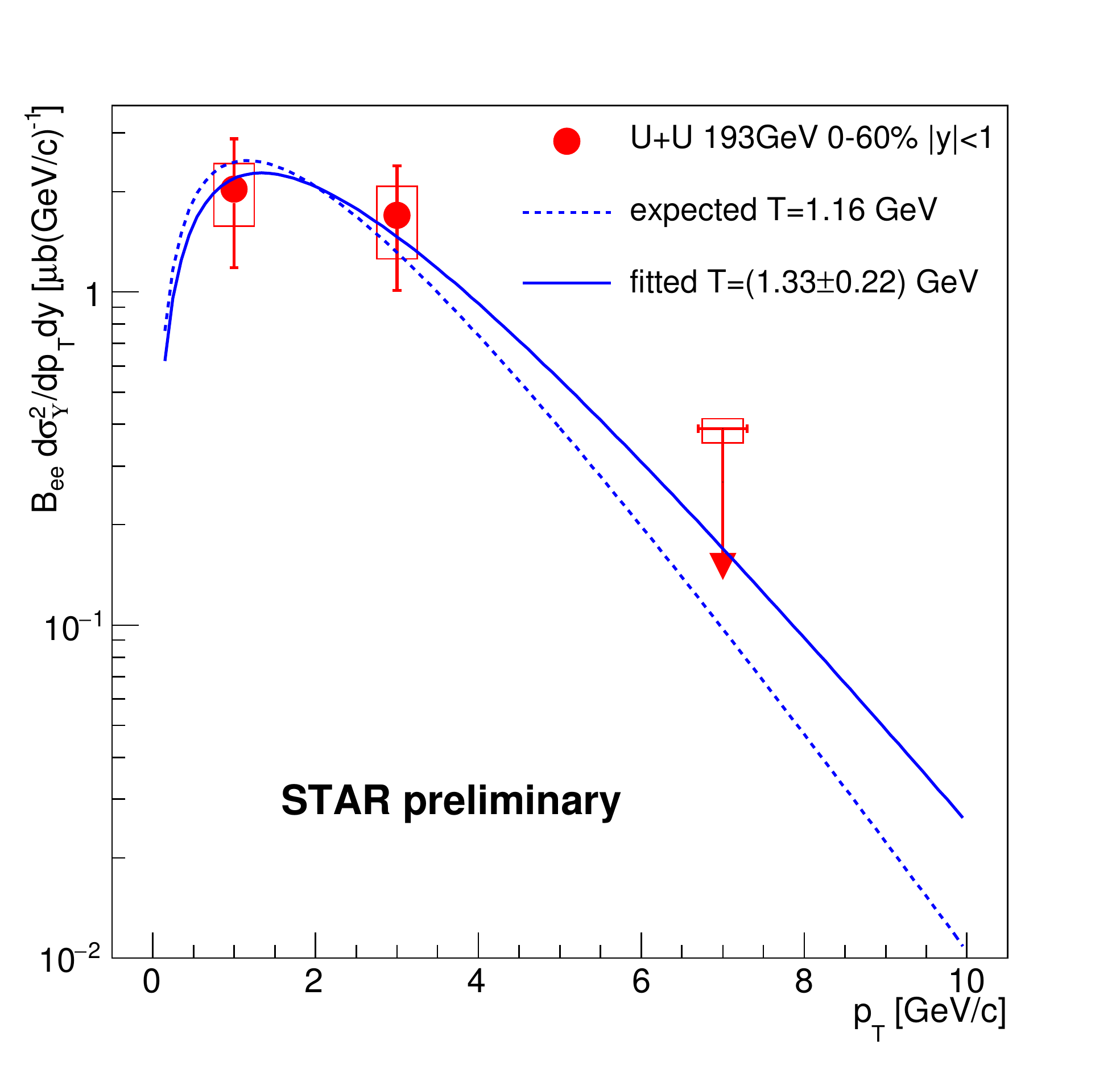}
\vspace{-3mm}
\caption{The \Ups{}(1S+2S+3S) \pT-spectrum in $\sqrt{s_{NN}}$=193 GeV U+U collisions.}
\label{fig:UpsilonSpectrum}
\end{figure}

The corrected \pT-spectrum of \Ups mesons produced in 0-60\% centrality U+U collisions at $\snn{}=193$ GeV is shown in Fig.~\ref{fig:UpsilonSpectrum}. Bin-shift correction was done using a Boltzmann function with a slope $T=1.16$ GeV, extracted from a parametrized interpolation over  p+p($\bar{\mathrm{p}}$) data from ISR, CDF and CMS. A fit to the spectrum yields a slope $T=1.32\pm0.21$ GeV. This is consistent with the interpolated p+p shape, in accordance with the flat \Raa(\pT) that was observed by CMS at LHC energies~\cite{HP15CMS}. 

\begin{figure}[]
\vspace{-2mm}
\includegraphics[width=0.95\linewidth]{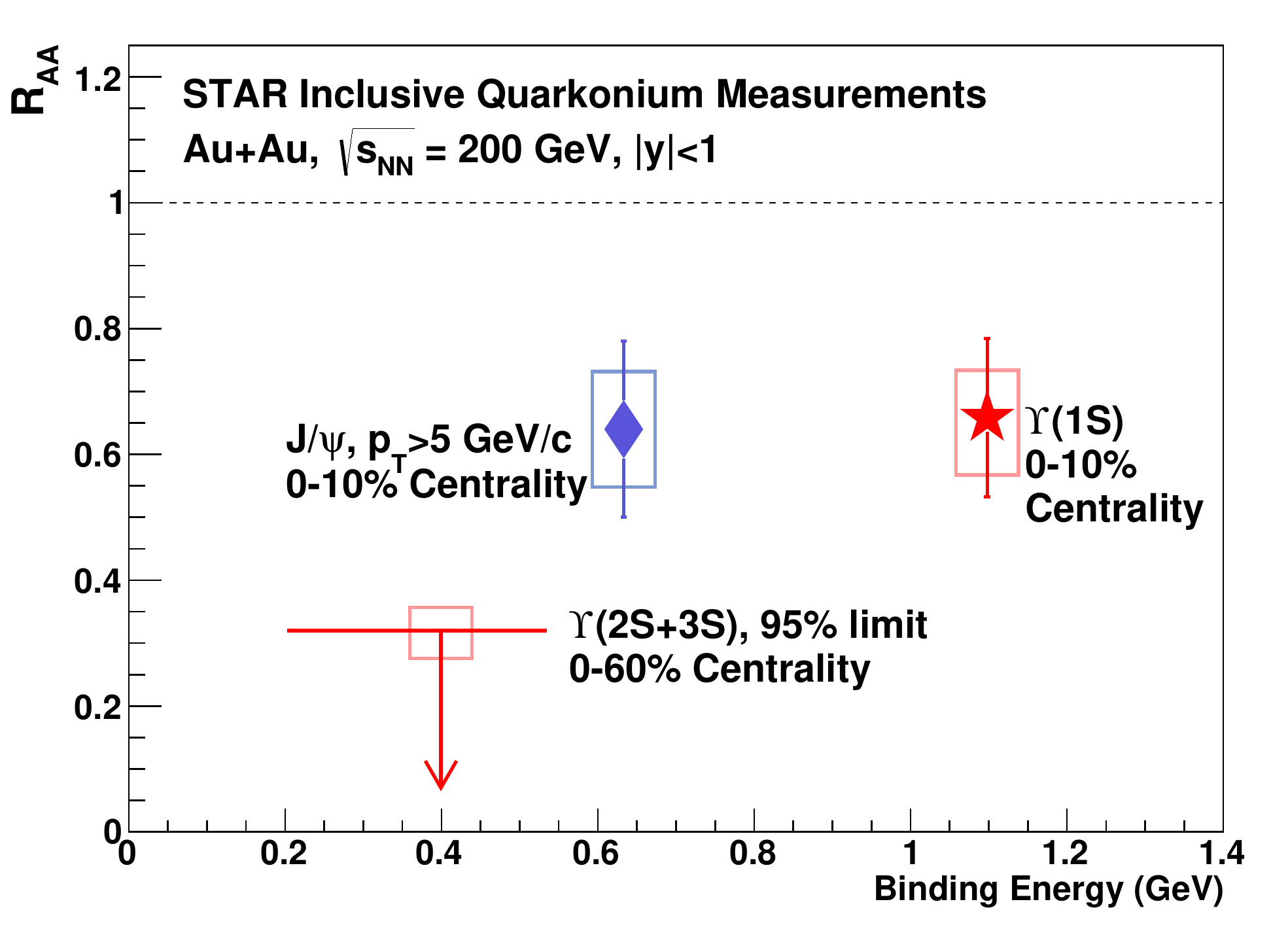}
\vspace{-3mm}
\caption{The \Raa of different quarkonium states versus binding energy in Au+Au collisions at $\sqrt{s_{NN}}$=200 GeV. The \Ups{}(1S) (star) and the high-\pT \Jpsi (diamond) points represent data of 0-10\% centrality, while the \Ups{}(2S+3S) 95\% upper limit (arrow) is of 0-60\% centrality.}
\label{fig:UpsRaaBinding}
\end{figure}

\begin{figure*}[]
\centering
\vspace{-2mm}
\includegraphics[trim={0 0 20mm 0},clip,height=60mm]{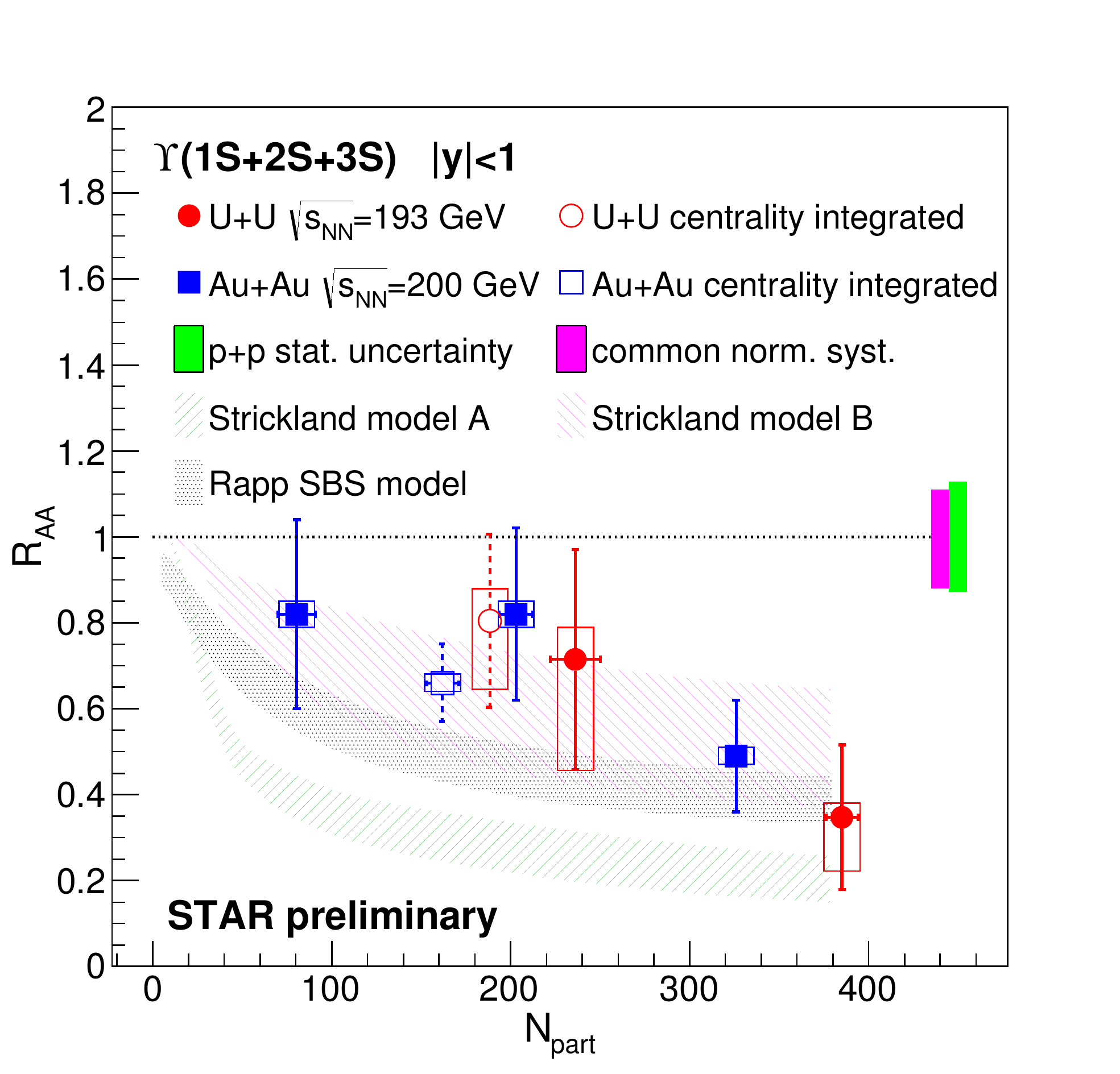}%
\includegraphics[trim={20mm 0 20mm 0},clip,height=60mm]{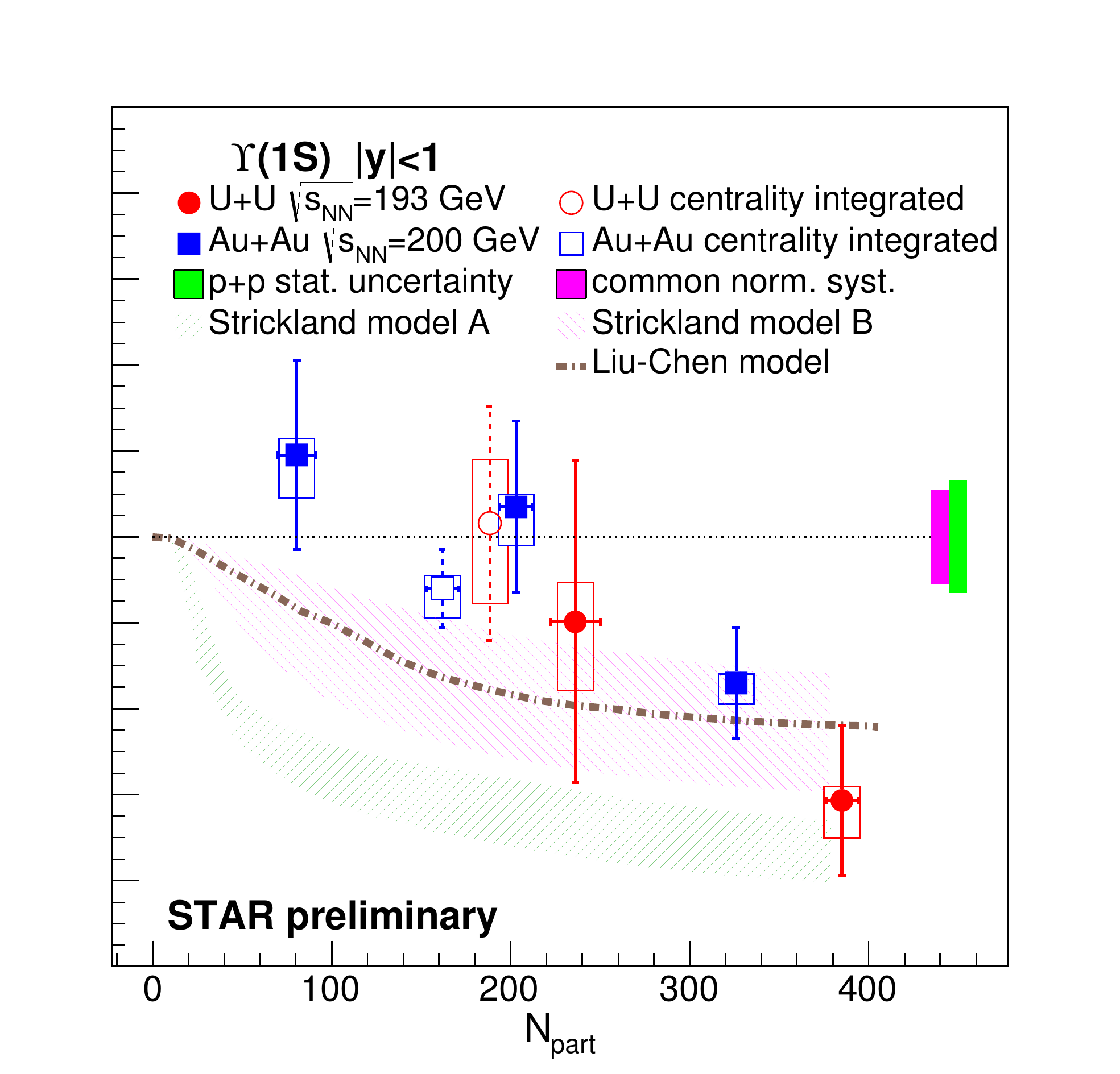}%
\includegraphics[trim={20mm 0 0 0},clip,height=60mm]{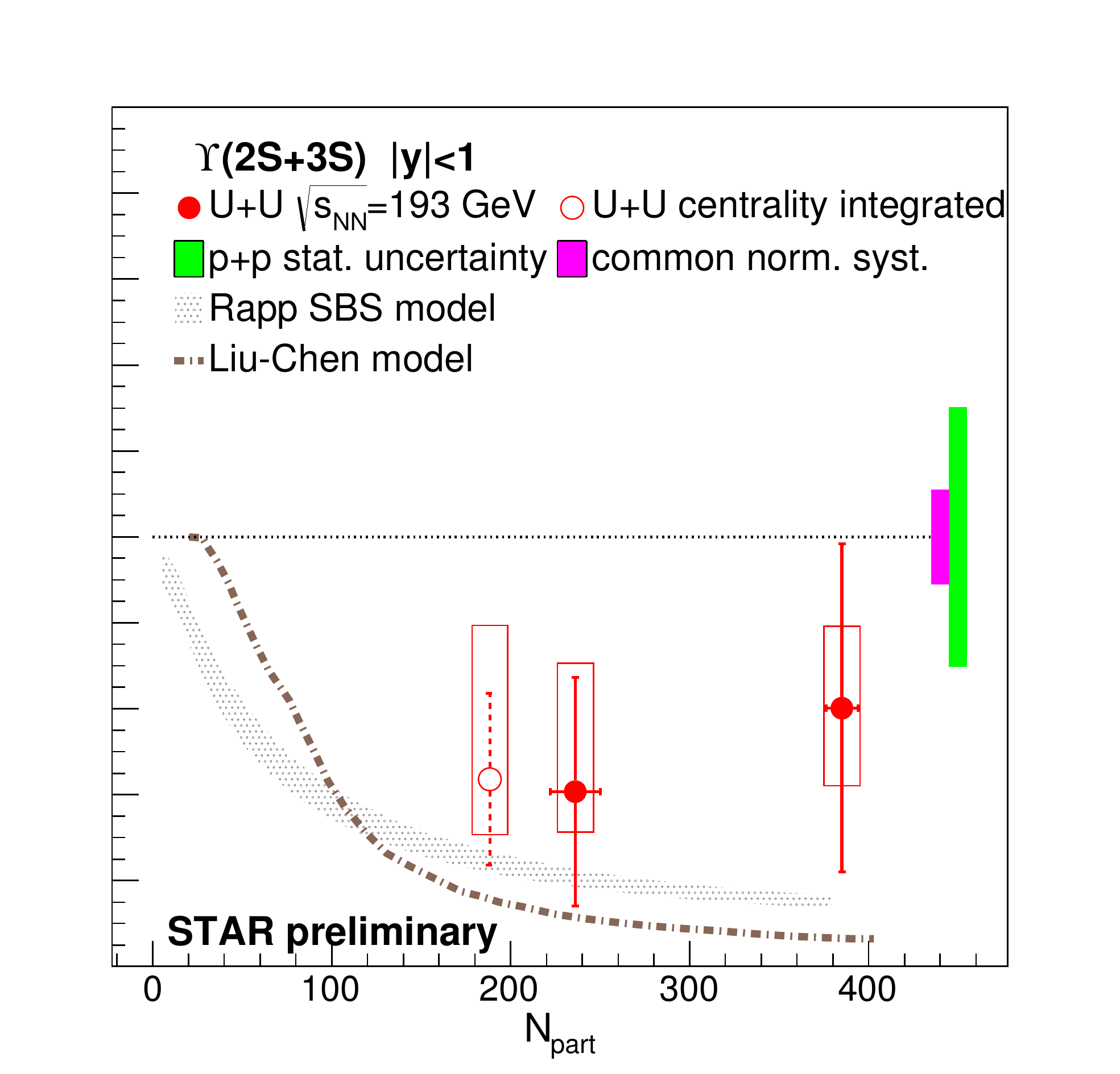}
\vspace{-3mm}
\caption{From left to right, \Raa of the \Ups{}(1S+2S+3S), \Ups{}(1S) and the \Ups{}(2S+3S) states plotted against \Npart, in $\sqrt{s_{NN}}$=200 GeV Au+Au (squares) and $\sqrt{s_{NN}}$=193 GeV U+U collisions (circles), compared to different models. Open symbols represent the 0-60\% centrality measurements. The models on the rightmost figure are \Ups{}(2S) only.}
\label{fig:UpsRaaNpart}
\end{figure*}

The \Raa of \Ups{}(1S) in central events, as well as the excited states \Ups{}(2S+3S) at 0-60\% centrality are compared to high-\pT \Jpsi mesons in $\sqrt{s_{NN}}=200$ GeV central Au+Au collisions~\cite{Adamczyk:2012ey} in Fig.~\ref{fig:UpsRaaBinding}. 
The nuclear modification factors of \Ups{}(1S+2S+3S), \Ups{}(1S) and \Ups{}(2S+3S) from Au+Au collisons at $\sqrt{s_{NN}}=200$ GeV as well as from U+U collisions at \snn{}=193 GeV are shown in Fig.~\ref{fig:UpsRaaNpart} with respect to \Npart, in comparison to theoretical calculations~\cite{Emerick:2011xu,Strickland:2011aa,Liu:2010ej}.

The trends observed in U+U data generally follow those in Au+Au and extends them towards higher \Npart values.
The emerging pattern of increasing suppression with higher \Npart and lower binding energy is consistent with the sequential melting hypothesis. 
The Au+Au $\Upsilon$(1S) shows a suppression similar to that of high-\pT \Jpsi mesons, more than if only cold nuclear matter effects were present~\cite{Adamczyk:2013poh}. Suppression of \Ups{}(1S) is significant, considering Au+Au and U+U results together. In 0-60\% centrality Au+Au collisions the excited state yields are consistent with a complete suppression within the precision of the measurement. Albeit the U+U results are consistent with  Au+Au, there is a hint with a significance of 1.8 $\sigma$ that $\Raa>0$ for the \Ups{}(2S+3S) states in U+U collisions of 0-60\% centrality. 
While the suppression of the \Ups states in central heavy-ion collisions at RHIC and in LHC~\cite{Chatrchyan:2012lxa} Pb+Pb collisions at \snn=2.76 TeV are comparable, RHIC data show a somewhat steeper \Npart dependence than their LHC counterparts. This may suggest that suppression is primarily driven by the energy density and not solely by the number of participant nucleons.

The model of Strickland and Bazow~\cite{Strickland:2011aa} incorporates lattice QCD results on bottomonium screening and thermal broadening, as well as the dynamical propagation of the $\Upsilon$ meson in the colored medium. Assuming an initial central temperature 428$<$T$<$444 MeV, the scenario with a potential based on heavy quark internal energy is consistent with the observations, while the free energy based scenario is disfavoured. 
The model of Liu {\it et al.}~\cite{Liu:2010ej} uses an internal-energy-based potential and an input temperature $T$=340 MeV.
The strong binding scenario in a model proposed by Emerick,
Zhao, and Rapp~\cite{Emerick:2011xu}, which includes possible CNM effects in
addition, is also consistent with STAR results. 

\section{Summary and outlook}

We observe a significant suppression of bottomonium states in central heavy-ion collisions. While the \Ups{}(1S) is similarly suppressed in $\sqrt{s_{NN}}$=200 GeV Au+Au collisions to the high-\pT \Jpsi mesons, there is a stronger suppression in the case of the excited states \Ups{}(2S+3S). 
This attests to the sequential melting picture in the presence of a deconfined medium. The suppression pattern seen in the case of the U+U collisions is consistent with the trend marked in Au+Au collisons and extends it towards higher \Npart values. 
The indication of \Ups{}(1S) suppression in central Au+Au collisions is confirmed by the new measurements in U+U. There is a hint with an 1.8 $\sigma$ significance that \Ups{}(2S+3S) is not completely suppressed in 0-60\% U+U collisions at $\sqrt{s_{NN}}$=193 GeV.

The precision of current measurements in the dielectron channel are limited by the low statistics of the heavy-ion and the reference samples.
Ongoing analysis of the high-statistics $\sqrt{s_{NN}}$=200 GeV Au+Au dataset, taken in 2014 with the recently inaugurated MTD detector, will strongly reduce both the statistical and systematic uncertainties. 
We expect separate \Raa measurements of all three \Ups{}(1S), \Ups{}(2S) and \Ups{}(3S) states with strongly reduced uncertainties~\cite{Yang:2014xta}. High statistics p+Au collisions taken in 2015 will
help us gain a deeper insight to the CNM effects.

This work has been supported by the grant 13-20841S of the Czech Science Foundation (GACR) and  by the MEYS grant CZ.1.07/2.3.00/20.0207 of the European Social Fund (ESF) in the Czech Republic: “Education for Competitiveness Operational Programme” (ECOP).

\end{document}